\documentclass[twocolumn,trackchanges]{aastex631}
\usepackage{rotating}
\usepackage{xcolor}
\usepackage{graphicx}
\usepackage{hyperref}
\usepackage{upgreek}

\date{Accepted XXX. Received YYY; in original form ZZZ}

\newcommand{\ubc}{Department of Physics \& Astronomy, University of British Columbia, 6224 Agricultural Road, Vancouver, BC V6T 1Z1, Canada}
\newcommand{\uconn}{Department of Physics, University of Connecticut, Storrs, CT, USA}
\newcommand{\cca}{Center for Computational Astrophysics, Flatiron Institute, New York, NY, USA}
\newcommand{\mcgilldep}{Department of Physics, McGill University, 3600 rue University, Montr\'eal, QC H3A 2T8, Canada}
\newcommand{\mcgillsi}{McGill Space Institute, McGill University, 3550 rue University, Montr\'eal, QC H3A 2A7, Canada}

\newcommand{\dunlap}{Dunlap Institute for Astronomy \& Astrophysics, University of Toronto, 50 St. George Street, Toronto, ON M5S 3H4, Canada}
\newcommand{\wvupa}{Department of Physics and Astronomy, West Virginia University, PO Box 6315, Morgantown, WV 26506, USA }
\newcommand{\wvugwac}{Center for Gravitational Waves and Cosmology, West Virginia University, Chestnut Ridge Research Building, Morgantown, WV 26505, USA}
\newcommand{\tifr}{Department of Astronomy and Astrophysics, Tata Institute of Fundamental Research, Mumbai, 400005, India}
\newcommand{\ncra}{National Centre for Radio Astrophysics, Post Bag 3, Ganeshkhind, Pune, 411007, India}
\newcommand{\UVA}{Anton Pannekoek Institute for Astronomy, University of Amsterdam, Science Park 904, 1098 XH Amsterdam, The Netherlands}
\newcommand{\curtin}{International Centre for Radio Astronomy Research, Curtin University, Bentley, WA 6102, Australia}

\newcommand{\cfrb}{CHIME/FRB}
\newcommand{\presto}{PRESTO}

\newcommand \rthree{FRB 20180916B}
\newcommand \rone{FRB 20121102A}
\newcommand \rseven{FRB 20190116A}
\newcommand \rsixteen{FRB 20190117A}

\newcommand{\rsixtyseven}{FRB 20201124A}

\newcommand \dmunits{$\mathrm{pc\,cm^{-3}}$}

\begin{document}

\title{Non-detection of \cfrb{} sources with the Arecibo Observatory}
\correspondingauthor{Deborah C. Good}
\email{deborah.good@nanograv.org}

\author[0000-0003-1884-348X]{Deborah C.~Good}
\affiliation{\uconn}\affiliation{ \cca}

\author[0000-0002-3426-7606]{Pragya Chawla}
\affiliation{\mcgilldep}\affiliation{\mcgillsi}\affiliation{\UVA}

\author[0000-0001-8384-5049]{Emmanuel Fonseca}
\affiliation{\wvupa} \affiliation{\wvugwac}
 \affiliation{\mcgilldep}
\affiliation{\mcgillsi}

\author[0000-0001-9345-0307]{Victoria Kaspi}
\affiliation{\mcgilldep}\affiliation{\mcgillsi}

\author[0000-0001-8845-1225]{B.~W.~Meyers}
\affiliation{\curtin} \affiliation{\ubc}

\author[0000-0002-4795-697X]{Ziggy Pleunis}
\affiliation{\dunlap}

\author[0000-0003-3154-3676]{Ketan R.~Sand}
\affiliation{\mcgilldep}\affiliation{\mcgillsi}

\author[0000-0002-7374-7119]{Paul Scholz}
\affiliation{\dunlap}

\author[0000-0001-9784-8670]{I.~H.~Stairs}
\affiliation{\ubc}

\author[0000-0003-2548-2926]{Shriharsh P.~Tendulkar}
\affiliation{\tifr}\affiliation{\ncra}

\begin{abstract}
    In this work, we present follow-up observations of two known repeating fast radio bursts (FRBs) and seven non-repeating FRBs with complex morphology discovered with \cfrb{}. These observations were conducted with the Arecibo Observatory 327 MHz receiver. We detected no additional bursts from these sources, nor did \cfrb{} detect any additional bursts from these sources during our follow-up program. Based on these non-detections, we provide constraints on the repetition rate, for all nine sources. We calculate repetition rates above 1 Jy using both a Poisson distribution of repetition and the Weibull distribution of repetition presented by \cite{oyp18}. For both distributions, we find repetition upper limits of the order $\lambda = 10^{-2} - 10^{-1} \text{hr}^{-1}$ for all sources. These rates are much lower than those published for notable repeating FRBs like \rone{} and \rsixtyseven, suggesting the possibility of a low-repetition sub-population.
\end{abstract}


\section{Introduction}
\label{sec:intro}

Fast radio bursts (FRBs) are luminous, millisecond transient signals detected at radio frequencies. FRBs have been studied since 2007 \citep{lbm+07}, and the published source list was expanded in 2021 by the addition of \cfrb{} Catalog 1, containing more than than 500 FRBs \citep{cfrbcatalog}. Though initially hard to detect in large quantities, FRBs are ubiquitous, with all-sky rates near 800 per sky per day above 5 Jy ms \citep{cfrbcatalog}. Though initially detected only as single events, the first FRB repetition was discovered in 2015 \citep{ssh+16a, ssh+16b}. It is now clear that many FRBs repeat, although the relationship between repeating and non-repeating FRB populations remains ambiguous. 

FRBs may all repeat with a variety of rates or there may be two source classes \citep{ravi19, csr19, cmg20}. Many repeating FRBs have been observed to burst only a few times, but others, like \rone{}, \rthree{}, and \rsixtyseven{} have been observed to burst prolifically \citep[see, e.g., ][]{RN,zww+21,r67}. Distribution of burst width and spectral properties in \cfrb{} Catalog 1 suggest the possibility of separate repeating and non-repeating populations \citep{cfrbcatalog, pgk+21}. Understanding the repetition behavior of a large sample of FRBs and understanding in detail the behavior of individual repeating FRBs may be a step towards understanding repeaters' place in the FRB ecosystem.

Among known repeaters, FRB repetition properties are an active area of study. Initially, bursts from \rone{} were observed to be both non-Poissonian and non-periodic \citep{oyp18}. In recent years, we have become aware of predictable periods of inactivity from both \rthree{} and \rone{} \citep{R3periodicity, rms+20,css+20}. Separately, \cite{millisecondperiodicity} found millisecond periodicity within the burst envelopes of at least one apparently non-repeating FRBs. 

Burst properties have also been of interest in studying repeating FRBs. Some of these bursts show complex downward-drifting spectro-temporal sub-structure, \citep[see, e.g., ][]{hss+19, R2, RN, RN2, pgk+21}. Additionally, many bursts show short timescale structure, including several detections of structure in the $10 \mu$s regime and one detection of $\lesssim 100$ ns  structure \citep{msh+18, nhk+21, mpp+21}. Many surveys, including \cfrb{} when not using baseband data, do not have sufficient time resolution to see such structure, and therefore follow-up observations with higher time resolution are needed.

There appear to be many more dim bursts from FRB sources than bright bursts, which could contribute to observations of complicated repetition properties among FRBs \citep{cp18}. Observations of FRB 20171019A (originally discovered by the Australian Square Kilometer Array Pathfinder Commensal Real-time ASKAP Fast Transients Survey, a.k.a., CRAFT) with the Green Bank Telescope revealed two repeat bursts a factor of about 600 fainter than the original ASKAP detection, lending strong support to the existence of a broad distribution of burst luminosities from repeating FRBs \citep{kso+19, zww+21}. This motivates further follow-up observations, as initial survey instruments may detect only the brightest bursts from repeating FRBs and detailed follow-up may allow us to fill in the luminosity distribution.

Studies of FRB repetition have thus far focused on a handful of highly prolific sources, such as \rone{} \citep[mean burst rate 27.8 per hour at 1.25 GHz during an active period; ][]{zww+21}, \rthree{} \citep[about 0.9 bursts per hour between 400-800 MHz during an active period; ][]{R3periodicity}, and \rsixtyseven{} \citep[about 10 bursts per day overall and about 100 bursts per day during a period of high activity; ][]{r67}. However, most of the repeating FRB population remains low-repetition rate bursts, observed only a few times. Therefore,  we must observe some low-repetition rate bursts in-depth to develop a complete picture of FRB repetition and burst morphology.

Though it has been very effective at expanding the known FRB source list, \cfrb{} has weaknesses. \cfrb{} intensity data has a time resolution of 0.983 ms, making it insensitive to short time-scale structure. The \cfrb{} baseband system has much higher time resolution, but baseband dumps are not available for all detections \citep{mmm+21}. \cfrb{} also has a relatively modest fluence sensitivity, reporting a median 95\% completeness across all bursts above a fluence of 5 Jy ms \citep{cfrbcatalog}. As a transit telescope located at latitude $49^{\circ}$ N, \cfrb{}'s exposure time is strongly declination dependent, and declinations near the telescope's southern limit (approximately $-20^{\circ}$) are observed for much shorter times than those further north. See \cite{chimefrb} for a more detailed discussion. Follow-up observations can provide focused observing time for under-observed declinations, higher time-resolution, and potentially sensitivity to dimmer bursts.

In this work, we report on follow-up observations of two low-declination repeating sources and seven low-declination non-repeating sources with Arecibo Observatory. Throughout this work, we characterize FRBs using their spectral flux density (Jy) and not spectral fluence density (Jy ms). In Section \ref{sect:obs_details}, we present information on the seven one-off and two repeating FRBs we observed as well as our observation program. In Section \ref{sect:sensitivity_limits} we set limits on observation sensitivity for \cfrb{} and Arecibo Observatory observations. In Section \ref{sect:burst_rate}, we present repetition rates and upper limits using Poisson errors and the Weibull distribution method of \cite{oyp18}. In Section \ref{sect:compare}, we compare our sources to other repeating FRBs discovered by \cfrb{}. Finally, in Section \ref{sect:discussion}, we discuss the implications of our results.

\section{Observation and Analysis Details \label{sect:obs_details}}
\subsection{Source information}
 We observed two sets of sources with Arecibo Observatory at 327 MHz, increasing our exposure time to these locations and providing sensitive, high time and frequency resolution observations. The first set was known repeaters at low declination, \rseven{} and \rsixteen{}, hereafter referred to as Group A. The primary goal of observing Group A sources was high time resolution observations of repetition and detection of low-luminosity bursts. The second class was seven one-off FRBs with the distinctive downward-drifting frequency structure often observed in repeating FRBs, hereafter referred to as Group B. The primary goal of observing Group B sources was to detect possible repetition. Burst properties are listed in Table \ref{tab:sourceinfo}.


\begin{sidewaystable}[hbtp]
    \begin{center} 
     \caption{A summary of burst properties, from \cite{RN, RN2, cfrbcatalog}}
    \label{tab:sourceinfo}
    \begin{tabular}{lcccccccc}
    \hline \hline
       Group & Source  & RA & Dec. & DM \tablenotemark{a}  & $N_{\text{bursts}}$ & Burst Width \tablenotemark{b}  & Scattering Timescale \tablenotemark{c} & Scattering Timescale  \\
       & & (hh:mm) & (dd:mm) & (pc/cm$^3$) & &  (ms) & (at 600 MHz; ms) &  (at 327 MHz; ms)  \\ \hline
        
     Group A &  \rseven{} & 12:49:19(35) & 27:09(14) & 445.5(2) & 2 & 2.8(4) & $<6.4$ & $<72$  \\ 
     Group A &   \rsixteen{} & 22:06:50(36) & 17:22(15)  & 393.360(2) & 6 & 3.6(4)  & 3.9(3) & 44(3) \\ 
	 Group B & FRB 20190109A &	07:11:50(59) & 	05:09(20) &  324.60(2) & 1 &   5(3) &  $<2.2$ & $<25$\\
	 Group B & FRB 20181030E &	09:02:41(3) &	08:53(11) & 159.690(4) &  1 &    $<0.39$  & 0.97(7) & 11.0(8)\\
	 Group B & FRB 20181125A &	09:51:46(43) & 	33:55(2) & 272.19(2) &   1 & 4.3(7) & $<1.5$ & $<18$\\
	 Group B & FRB 20181226B & 12:10:38(9) &	12:25(11) & 287.036(2) &  1 &  2.5(2)  & 1.10(4) & 12.4(5) \\
	 Group B & FRB 20190124C &	14:29:41(11) & 	28:23(10) & 303.644(1) &  1 &   2.06(9) & 0.67(2) & 7.6(3)\\
	 Group B & FRB 20190111A &	14:28:01(37) &  26:47(7) & 171.9682(1) &  1 &  1.58(4)  & 0.53(3)  & 6.1(3) \\
	 Group B & FRB 20181224E &	15:57:17(8) & 	07:19(12) &  581.853(4) & 1 &  2.1(2) & $<1.1$ & $<13$\\ \hline
    \end{tabular}
    \end{center}
    \tablenotetext{a}{DMs for Group A sources are averages of the values reported in \cite{RN} and \cite{RN2}, and reflect the uncertainty in measurements, but not the range of observed values. DMs for Group B sources are those reported in \cite{cfrbcatalog}.}
    \tablenotetext{b}{Widths are boxcar widths determined in \cite{cfrbcatalog}. For bursts with multiple components, components have been summed to determine total width. For Group A sources, average values are reported, and uncertainties are based on measurement uncertainty, not the range of measured values.}
    \tablenotetext{c}{Scattering time at 600 MHz. Where only an upper limit is available, we have used it as $t_{\text{scatt}}$ For Group A sources, average values are reported, and uncertainties are based on measurement uncertainty, not the range of measured values.}
  \end{sidewaystable}

All Group A and B sources are included in \cfrb{} Catalog  1, and dynamic spectra as well as detailed information can be found in \cite{cfrbcatalog}. Group A sources are also included in \cite{RN} or \cite{RN2}. We began this project observing \rseven{} and Group B sources, but later added \rsixteen{} and stopped observing Group B sources. 

In Table \ref{tab:exposure}, we include the total exposure time with \cfrb{}, as well as with Arecibo Observatory. All exposures for \cfrb{} are for the period between 2018-08-28 (MJD 58358) to 2021-05-01 (MJD 59335). Arecibo Observatory observations for Group A source were conducted between 2019-08-17 (MJD 58712) and 2020-06-10 (MJD 59010). Arecibo Observatory observations for Group B sources were conducted between 2019-07-31 (MJD 58695) and 2019-12-31 (MJD 58848).

\begin{table*}[hbtp]
    \begin{center} 
     \caption{A summary of exposure with \cfrb{} and AO for all sources}
    \label{tab:exposure}
    \begin{tabular}{lccc}
    \hline \hline
        Source  &  \cfrb{}  & AO & AO Obs. Dates \\
        & exposure (hr)\tablenotemark{a} & exposure (hr) \tablenotemark{b} & (MJD) \\ \hline
     \rseven{}\tablenotemark{c} & $45.8 \pm 27.1$ & 27.1 & 58712 -- 59009   \\ 
    \rsixteen{}\tablenotemark{c} & $44.7 \pm 24.4$  & 20.1   & 58888 -- 59010\\ 
	FRB 20190109A &	    $40.2 \pm 23.5$ & 4.78 & 58719 --58822 \\
	FRB 20181030E &	   $50.1 \pm 17.8$  & 4.00 & 58754 -- 58822\\
	FRB 20181125A &	   $32.5 \pm 31.4$ & 2.90 & 5874 -- 58822\\
	FRB 20181226B &    $52.2 \pm 17.0$  & 4.85 & 58712-- 58848\\
	FRB 20190124C &    $50.3 \pm 20.1$ & 2.12 & 58754 -- 58824\\
	FRB 20190111A &    $57.5 \pm 16.8$  & 3.07 & 58754 -- 58832  \\
	FRB 20181224E &	  $52.2  \pm 12.5$ & 2.69 & 58701 -- 58832\\ \hline
    \end{tabular} \\
    \end{center}
    \tablenotetext{a}{Exposure uncertainties are set by \cfrb{} position uncertainties. All \cfrb{} exposures are for the period between August 28, 2018 (MJD 58358) and May 1, 2021 (MJD 59335).}
   \tablenotetext{b}{Arecibo Observatory exposures are uncorrected for efficiency losses}
    \tablenotetext{c}{Known repeater.}
  \end{table*}

\subsection{Search Strategy}
We conducted observations using the Arecibo Observatory 327 MHz receiver and the Puerto Rico Ultimate Pulsar Processing Instrument (PUPPI) backend. Best-fit positional uncertainties from \cfrb{} are of the order of 10$^{\prime}$. The 327 MHz receiver operates at a complementary frequency range to that of \cfrb{} and has a comparable beam size, which allowed us to avoid gridded observations. See comparison of receiver properties in Table \ref{tab:ao_chime_props}.

\begin{table}
\begin{center}
\caption{Basic properties of Arecibo Observatory 327 MHz Receiver and of \cfrb{} \label{tab:ao_chime_props}}
\begin{tabular}{lcc} \hline \hline
& AO 327 & \cfrb{} \\ \hline
Beam Width  & 4$^{\prime}$ $\times$ 15$^{\prime}$ & 40$^{\prime}$ -- 20$^{\prime}$ \tablenotemark{a}    \\ 
Centre Frequency (MHz) & 327 & 600 \\ 
Bandwidth (MHz) & 50 & 400 \\ 
System Temperature (K) &115 & $\sim 50$ \\
Antenna Gain (K/Jy) & 10  & 1.16 \\ \hline
\end{tabular}
\\
\end{center}
\tablenotetext{a}{Due to \cfrb{}'s large fractional bandwidth, the beam width changes by about a factor of two over the band.}
\end{table}

We observed in coherently de-dispersed search mode, taking advantage of the known dispersion measures for each of our sources. DM has been observed to change between bursts from repeating FRB sources, so we anticipated the possibility that the nominal DM will not represent the best-fit DM for repeat bursts \citep{jcf+19, hss+19, hmsw+21, lwz+21}.  As we hoped to study bursts in high resolution, we recorded data and 10 $\mu$s time resolution and downsampled to 80 $\mu$s for initial analysis, providing about a factor of ten higher time resolution than \cfrb{} intensity data. Though we searched intensity data, we recorded full Stokes polarization information.

\subsection{Data Analysis}
Data were analyzed using standard single pulse search methods in \presto{} \citep{presto}. We began by searching data for RFI using \presto{}'s \texttt{rfifind} method. We downsampled the data in time and frequency to accelerate analysis. We determined an optimal dedispersion plan with \presto{}'s \texttt{DDplan.py} tool and followed it to search for bursts up to 30 \dmunits{} above and below the nominal DM of our source. Finally, we ran \texttt{single\_pulse\_search} to find signals from single pulses. Anything which appeared in the \texttt{single\_pulse\_search} with $S/N > 6$, we then visually inspected. Though a number of possible candidate events passed the $S/N > 6$ threshold, each was clearly non-astrophysical RFI when examined with \presto{}'s \texttt{waterfaller} functionality. 

\section{Results}
During the course of this project, we did not observe repeat bursts from any of the sources. Not only did we not observe any bursts with Arecibo Observatory, no further bursts from these sources were observed by \cfrb{} while the source was being observed with Arecibo Observatory. This means that none of our seven Group B sources has been observed to repeat during the duration considered, up to May 1, 2021 (MJD 59335). This also suggests genuinely low burst rates from Group A sources, particularly \rseven{}, which has not been observed to burst since its initial discovery in January 2019. \rsixteen{} has been observed to burst several times, but all bursts were prior to the start of follow-up observations.

In the remainder of this work, we will set upper limits on the repetition rate of each source. We will also discuss possible explanations for the non-detection of further repeat bursts and the possible implications of this low-repetition rate within the context of the broader FRB repetition discussion. 

Before discussing results in detail and setting these upper limits, it is important to note that using the first burst of a repeating FRB to estimate burst rate is inherently biased. If a population of bursts with extremely low burst rate (e.g., $\lambda = 10^{-9}$ years) exists, some examples of that burst population would be discovered, but additional bursts would not be detected during a human lifespan. Calculating a burst rate upper limit based on a short observation duration such as ours would result in a large over-estimate of burst rate. This is a systemic problem in discussing FRB repetition, and an attempt to solve it is beyond the scope of this work. The problem is discussed in more detail in \cite{james19}.

\subsection{Understanding Sensitivity Limits \label{sect:sensitivity_limits}}
In order to fully understand the implications of our non-detection, we set a sensitivity limit for our survey. To do this, we follow the extension to the radiometer equation derived by \cite{cm+03} which asserts that the minimum detectable flux density for a radio transient with a given receiver is 
\begin{equation}
S_{\text{min}} = \frac{\beta \text{S/N} \left(T_{\text{rec}} + T_{\text{sky}} \right)}{G W_i} \sqrt{\frac{W_b}{n_p \Delta \nu}},
\end{equation}
where $\beta$ is an instrumental factor accounting for digitization loss (approximately equal to 1), $S/N$ is signal-to-noise ratio, $T_{\text{rec}}$ is receiver temperature, $T_{\text{sky}}$ is average sky temperature, $G$ is antenna gain, $W_i$ is intrinsic pulse width, $W_b$ is broadened pulse width, $n_p$ is number of polarizations and $\Delta \nu$ is frequency bandwidth.

The broadened width, $W_b$ is the quadrature sum of the pulse's intrinsic width, the sampling time $t_{\text{samp}}$, the scattering time $t_{\text{scatt}}$, and the per-channel dispersive delay $t_{\text{chan}}$
\begin{equation}
W_b = \sqrt{W_i^2 + t_{\text{samp}}^2 + t_{\text{chan}}^2 + t_{\text{scatt}}^2}.
\end{equation}

The dispersive delay $t_{\text{chan}}$ as derived in e.g, \cite{lk04} is 
\begin{equation} 
t_{\text{chan}} = 8.3 \mu s \left(\frac{\Delta \nu_{\text{chan}}}{\text{MHz}} \right) \left( \frac{\nu}{\text{GHz}} \right)^{-3} \left(\frac{\text{DM}}{\text{pc cm}^{-3}}  \right),
\end{equation}
where $\Delta \nu$ is the frequency channel bandwidth and $\nu$ is the central observing frequency.

For known repeating bursts, \rseven{} and \rsixteen{}, we have used measured width parameters and scattering parameters reported in \cite{RN} and \cite{RN2}; for the potential repeaters, we have used the measured \cfrb{} parameters from \cite{cfrbcatalog}. FRBs have been observed to vary widely in repeat bursts, indeed even within the Group A sources. Therefore, we take averages for Group A parameters and treat all sensitivity estimates as educated approximations, not definitive values. As the sensitivity limits are predominantly used to compare the two observatories, the most important aspect is that the burst parameters are identical for both observatory's sensitivity calculations.

We calculate these limits for both \cfrb{} and Arecibo Observatory, to enable comparison of the sensitivities and to allow us to reliably combine data from the two observatories. The parameters used to calculate these results are shown in Table \ref{tab:sourceinfo} and \ref{tab:ao_chime_props}; the results are shown in Table \ref{tab:sensitivity_limits}. 

\begin{table}[hbtp]
\begin{center}
\caption{Sensitivity thresholds for Arecibo Observatory and \cfrb{}. \label{tab:sensitivity_limits}}
\begin{tabular}{lcc} 
\hline \hline \\ 
Source Name & AO 327 $S_{\text{min}}$\tablenotemark{a} & \cfrb{} $S_{\text{min}}$ \\
& Jy (327 MHz) & Jy (600 MHz) \\ \hline
    FRB 20190116A\tablenotemark{b}  & $0.84\pm 0.12 $  & $1.0 \pm 0.15 $  \\ 
    FRB 20190117A\tablenotemark{b}  & $0.51 \pm 0.068 $ & $0.88 \pm 0.089$ \\
	FRB 20190109A & $0.29 \pm 0.16$ & $0.50 \pm 0.35$	  \\
	FRB 20181030E & $2.5 \pm 0.11$ & $3.2\pm 0.10 $	  \\
	FRB 20181125A & $0.30	\pm 0.049$ & $0.53 \pm 0.12 $	\\
	FRB 20181226B & $0.48	\pm 0.045$ & $0.73 \pm0.084$	  \\
	FRB 20190124C & $0.55\pm 0.025 $	& $0.80 \pm 0.046$	\\
	FRB 20190111A & $0.56 \pm 0.018$	& $0.93 \pm 0.032 $	 \\
	FRB 20181224E & $0.74 \pm 0.076 $	 & $0.82 \pm 0.10 $ 	 \\ \hline
\end{tabular} \\
\end{center}
\tablenotetext{a}{See Table \ref{tab:ao_chime_props} for parameters. Sensitivity is calculated using the formalism of \cite{cm+03}}
\tablenotetext{b}{Known repeater}
\end{table}

The Arecibo Observatory sensitivity thresholds are lower than the \cfrb{} sensitivity limits, though not by as large a factor as a simple radiometer equation calculation might suggest. This is largely due to the effects of scattering. Scattering timescales for \cfrb{} detections are measured relative to 600\,MHz. To determine the scattering timescales for our Arecibo Observatory observations, we have used the standard scaling used for pulsar scattering, $\tau_s \propto \nu^{-4}$ \citep{rickett69}. Scaling to 327 MHz, therefore, increases the scattering timescales from those measured at 600 MHz, limiting the increase in sensitivity we are able to achieve by observing with Arecibo Observatory. Additionally, the 327 MHz receiver was among Arecibo Observatory's less sensitive receivers. However, as scattering timescales can vary between bursts for repeating FRBs, this can be regarded as a slightly pessimistic sensitivity measurement.

The sensitivity limit for FRB 20181030E is much higher than for the other bursts. This is because FRB 20181030E is much narrower than other bursts. Though in its initial detection, it appeared morphologically complex, subsequent analysis for \cfrb{} Catalog 1 has revealed that it is in fact a moderately narrow burst with structure likely better described by scattering than by sub-burst drifting. 

\subsection{Constraints on Burst Rate \label{sect:burst_rate}}
As we did not detect additional bursts from any of our sources, we focus on providing constraints on burst rate. It is not definitively known what the best distribution is to use for predicting FRB burst rates. With that in mind, we approach this problem from two directions. First, and simplest, we calculate limits on burst rate based on a Poisson distribution. However, it has been well established that a pure Poisson distribution does not well describe FRB emission. Therefore, we also use the Weibull distribution approach outlined by \cite{oyp18} and applied by other work in the field such as \cite{csr+19}. The Weibull distribution is itself an extension of the Poisson distribution with the addition of a clustering parameter. 

\subsubsection{Poisson distribution \label{sect:poisson}}

For a Poisson process with burst rate $r$, the distribution of intervals $\delta$ between bursts is exponential
\begin{equation}
P(\delta | r)  = r \, e^{-\delta \, r}.
\end{equation}
Though not the most physically motivated distribution for burst repetition, the Poisson distribution is well-understood and makes minimal assumptions about our observation program.

\begin{figure*}
\centering
\includegraphics[width=0.7\textwidth]{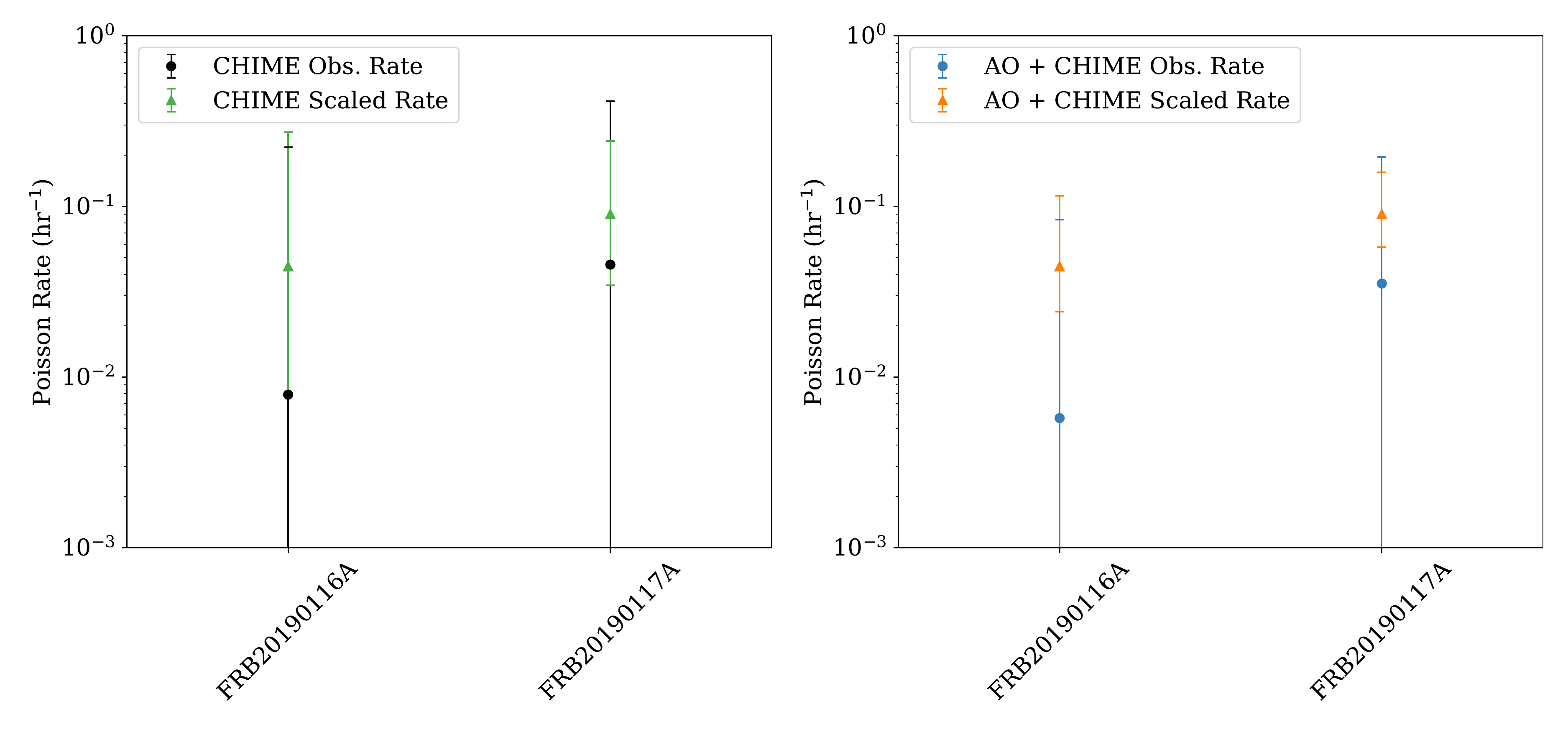}
\includegraphics[width=0.7\textwidth]{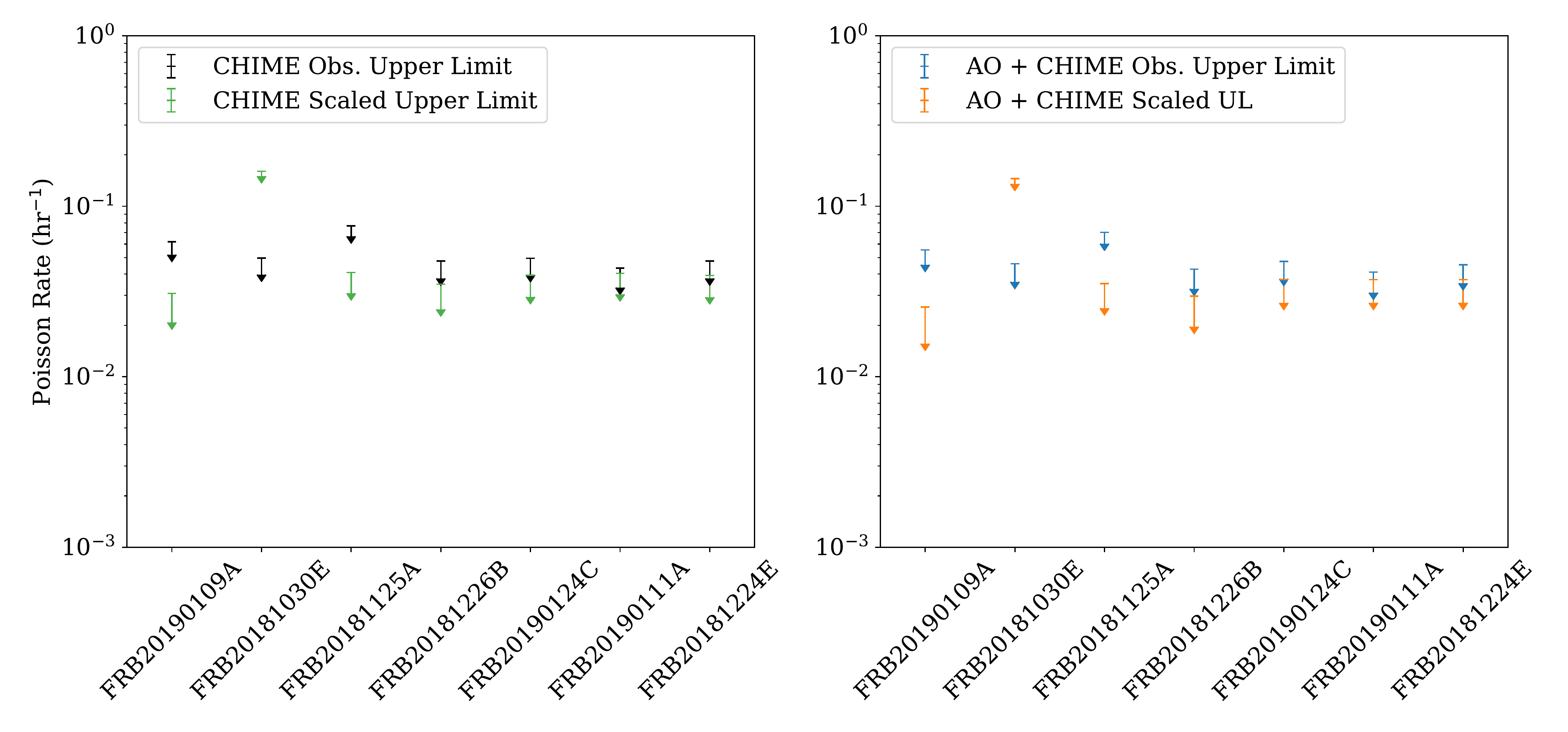}
\caption{Poisson repetition rates above 1 Jy for known repeaters (top panels) and upper limits on repetition for non-repeating bursts (bottom panels). In both cases, the left-hand panel shows observed and sensitivity scaled \cfrb{} only rates or limits, while the right-hand panel shows combined \cfrb{} and Arecibo Observatory rates and limits. Note that FRB 20181030E is much narrower than other bursts and therefore has a much higher sensitivity threshold. This causes the scaled rate to increase rather than decrease.  \label{fig:poisson_rates}}
\end{figure*}

To determine our repetition rate, we follow the example of \cite{RN} and present both observed rates and scaled rates, both above 1 Jy. All of our rates have Poisson confidence intervals, calculated as outlined in \cite{kbn91}, and as implemented in \texttt{astropy.stats}. We calculate rates for \cfrb{}-only and for \cfrb{}-Arecibo Observatory combined data. We do not calculate Arecibo Observatory-only rates, as these would be trivially zero. The exposure time and uncertainty in exposure time for \cfrb{} bursts is calculated as described in \cite{cfrbcatalog}. Exposure time for Arecibo Observatory is the sum of individual observations' duration for each source. These results are shown in Table \ref{tab:poisson_rep_rate} and Figure \ref{fig:poisson_rates}. As we would expect, the addition of significant Arecibo Observatory exposure without a detection decreases the burst rates. 

The scaled rate can be determined by multiplying the observed rate by $(S/S_0)$, where $S$ is our sensitivity limit in Jy and $S_0 = 1$ Jy. Instead of scaling the final rate, we instead scale the exposure time for each telescope, so that we can add the exposure times and calculate a final combined, scaled rate:

\begin{equation}
r_{\text{scaled}} = \frac{N_{\text{bursts}}}{T_{\text{C}} \left(S_{\text{C}}/S_0\right)^{1.0} +T_{\text{AO}} \left(S_{\text{AO}}/S_0\right)^{1.0} }.
\end{equation}

\begin{table*}[hbtp]
\begin{center}
\caption{Poisson repetition rates \tablenotemark{a} and upper limits for FRBs selected for followup with Arecibo Observatory\label{tab:poisson_rep_rate}}
\begin{tabular}{ccccc}
Source & \cfrb{}. & \cfrb{} & \cfrb{} + AO  & \cfrb{} + AO  \\
 & Obs & Scaled & Obs & Scaled  \\
 & (bursts/hr) & (bursts/hr) & (bursts/hr)  & (bursts/hr) \\   \hline \hline
FRB 20190116A\tablenotemark{b} & $0.044^{+0.259}_{0.008}$ &  $0.045^{+0.272}_{-0.008}$  & $0.027^{+0.105}_{-0.006}$ & $0.097_{-0.006}^{+0.097}$ \\ \\
FRB 20190117A\tablenotemark{b}  & $0.134^{+0.502}_{-0.045}$ &  $0.090^{+0.241}_{-0.034}$  & $0.093^{+0.252}_{-0.035}$  & $0.057_{-0.024}^{+0.125}$\\  \\
FRB 20190109A  & $< 0.062$ & $<0.031$ & $<0.055$ & $<0.026$ \\ \\
FRB 20181030E & $<0.050$ & $<0.16$ & $0.046$ & $<0.15$ \\  \\
FRB 20181125A  & $<0.077$  & $<0.041$ & $<0.070$ & $<0.035$ \\  \\ 
FRB 20181226B  & $<0.048$ & $<0.035$ & $<0.043$ & $<0.030$\\ \\
FRB 20190124C  & $<0.050$ & $<0.039$ & $<0.048$ & $<0.037$\\ \\
FRB 20190111A  & $<0.043$  & $<0.040$ & $<0.041$ & $<0.037$\\  \\
FRB 20181224E  & $<0.048$  & $<0.039$ & $<0.045$  & $<0.037$\\ \\
\hline 
\end{tabular}
\end{center}
\tablenotetext{a}{Above 1 Jy}
\tablenotetext{b}{Known repeater}
\end{table*}

\subsubsection{Weibull Distribution}

We follow closely the framework outlined by \cite{oyp18}, creating a Weibull distribution with parameters $r$ (rate) and $k$ (clustering parameter), determining Bayesian probabilities for these data, sampling over a grid of $r$ and $k$ values. The likelihood functions for Weibull-distributed repeater bursts are derived in detail in that work. The result is that the Poisson distribution is generalized such that the interval between bursts is
\begin{equation}
W(\delta|r, k) = k \,\delta^{-1} [\delta\, r\, \Gamma\left(1+1/k\right)]^k e^{- [\delta r \Gamma\left(1+1/k\right)]^k},
\end{equation}
where $\delta$ is the interval between bursts and $\Gamma$ is the incomplete $\Gamma$ function \citep{oyp18}. If $k = 1$, the Weibull distribution reduces to the Poisson distribution. In cases where $k<1$, the presence of one burst makes it more likely another burst will be detected. This method has been used extensively by the community since its introduction including in systematic efforts to understand FRB repetition such as \cite{csr19}, \cite{cp18}, and \cite{jof+20a}. 

The likelihood of a particular value of $k$ and $r$ is 
\begin{equation}
p(k,r | N, t_1...t_N) \propto p(N, t_1...t_N |  k,r) p(k,r), 
\end{equation}
where we use a Jeffrey's prior for $p(k,r)$, uniform sampling in $\log{k}$ and $\log{r}$, as in \cite{oyp18}, allowing us to generate values for $r$ and $k$ by creating a grid of values in $\log{(k)}$ and $\log{(r)}$. 

Under the assumption that gaps between observations are long relative to the length of observations, we can treat each individual observation as an independent event and can determine the total probability of our observed results by multiplying together the probability of individual observation outcomes. There are three possible outcomes for each observation: no bursts are detected ($N=0$), one burst is detected ($N=1$), or two or more bursts are detected ($N>1$). The vast majority of our observations are in the first category: no detection. Each Group B source has one single-detection observation (category 2) and Group A source \rsixteen{} has several single-detection (category 2) observations. The only observation in the third category, multiple detections in a single observation, is Group A's \rseven{} discovery observation with CHIME. All other observations of \rseven{} are non detections.

In general, the probability density $\mathcal{P}$ for $N$ bursts during a single observation of known duration, given $k$ and $r$ is 
\begin{eqnarray}
\label{eqn:prob_two_or_more}
\mathcal{P}(N, t_1...t_N |  k,r) = r\; \text{CDF}(t_1 | k, r)\;  \nonumber \\ \text{CDF}(\Delta - t_N | k, r)\;  \nonumber \\
\times \prod\limits_{i=1}^{N-1} (t_{i+1} - t_i | k,r),
\end{eqnarray}
where the cumulative distribution function (CDF) is $\text{CDF} = \exp{\left(-[\delta r \Gamma(1 + 1/k) ]^k\right)}$.

Equation \ref{eqn:prob_two_or_more} is strictly valid only for the case where more than one burst is detected in a single observation, which was the primary focus of the original work and is a common use case for the framework. However, \cite{oyp18} also derived special cases for $N=0$ and $N=1$, which we employ extensively here. The probability density for an observation of length $\Delta$ with one burst detected at time $t_1$ is
\begin{eqnarray} \label{eqn:prob_one}
\mathcal{P} \left(N=1 | k,r\right) = r \; \text{CDF}(t_1 | k, r)\; \nonumber \\\text{CDF}(\Delta - t_N | k, r).
\end{eqnarray}

The probability of an observation of length $\delta$ with no bursts detected during the observation is simply Equation \ref{eqn:prob_one} marginalized over the time of the first burst, i.e., a no-burst observation is simply a one-burst observation with $t_1 = \infty$
\begin{equation}
 \label{eqn:prob_zero}
P\left(N=0 | k,r \right) = \int_{\Delta}^{\infty} d t_1 \; \mathcal{P} (t_1 | k,r)
\end{equation}
and thus 
\begin{equation}
    P\left(N=0 | k,r \right) = \frac{\Gamma_i\left(1/k, (\Delta \; r \; \Gamma(1+1/k))^k\right)}{k\; \Gamma \left(1+1/k\right)}
\end{equation}
where $\Gamma_i$ is the incomplete $\Gamma$ function.

The total posterior for a set of observations is
\begin{equation}
    \mathcal{P}(k, r | N, t_1\ldots t_N) \propto \mathcal{P}(N, t_1 \ldots t_N | k, r)\; \mathcal{P} (k, r).
\end{equation}

This implementation was created with single dish observatories like Arecibo Observatory and the Green Bank Telescope in mind, and therefore it assumes observation duration and elapsed time from the start of the observation to the burst are clearly defined. However, due to \cfrb{}'s complex beam structure, the exact start and end times of \cfrb{} observations, and therefore the elapsed times between the burst and the start of the observation, are somewhat ambiguous. The exposure time used in other calculations in this work and in previous \cfrb{} publications is an attempt to quantify the amount of time a source is visible, but it does not account for the substantial daily time where sources are visible in the side-lobes of the beam. (See \citealt{chimefrb} and \citealt{cfrbcatalog} for more detailed discussion of the \cfrb{} system.) Between August 28, 2018 and May 1, 2021, there are approximately 760 days of observations that are deemed ``good''' for science analysis by the \cfrb{} exposure and sensitivity team. These observations are approximately three minutes each. The time an ``observation'' begins (i.e., when the location of the burst becomes visible to \cfrb{}) is frequency-dependent and complicated by side-lobe structure. Therefore, for simplicity's sake, 
we assume that the first burst to occur during an individual daily observation occurred at $t=0$ within that observation, i.e., no elapsed time between the start of the observation and the burst. We maintain the spacing within the observation between the first and second \rseven{} bursts. For the majority of observations, where no burst is detected, we use Equation \ref{eqn:prob_zero} and do not make any assumptions about prior bursts. To meaningfully combine \cfrb{} and Arecibo Observatory data, we use the scaling discussed in Section \ref{sect:sensitivity_limits}.

\begin{figure}
\centering
\includegraphics[width=0.4\textwidth]{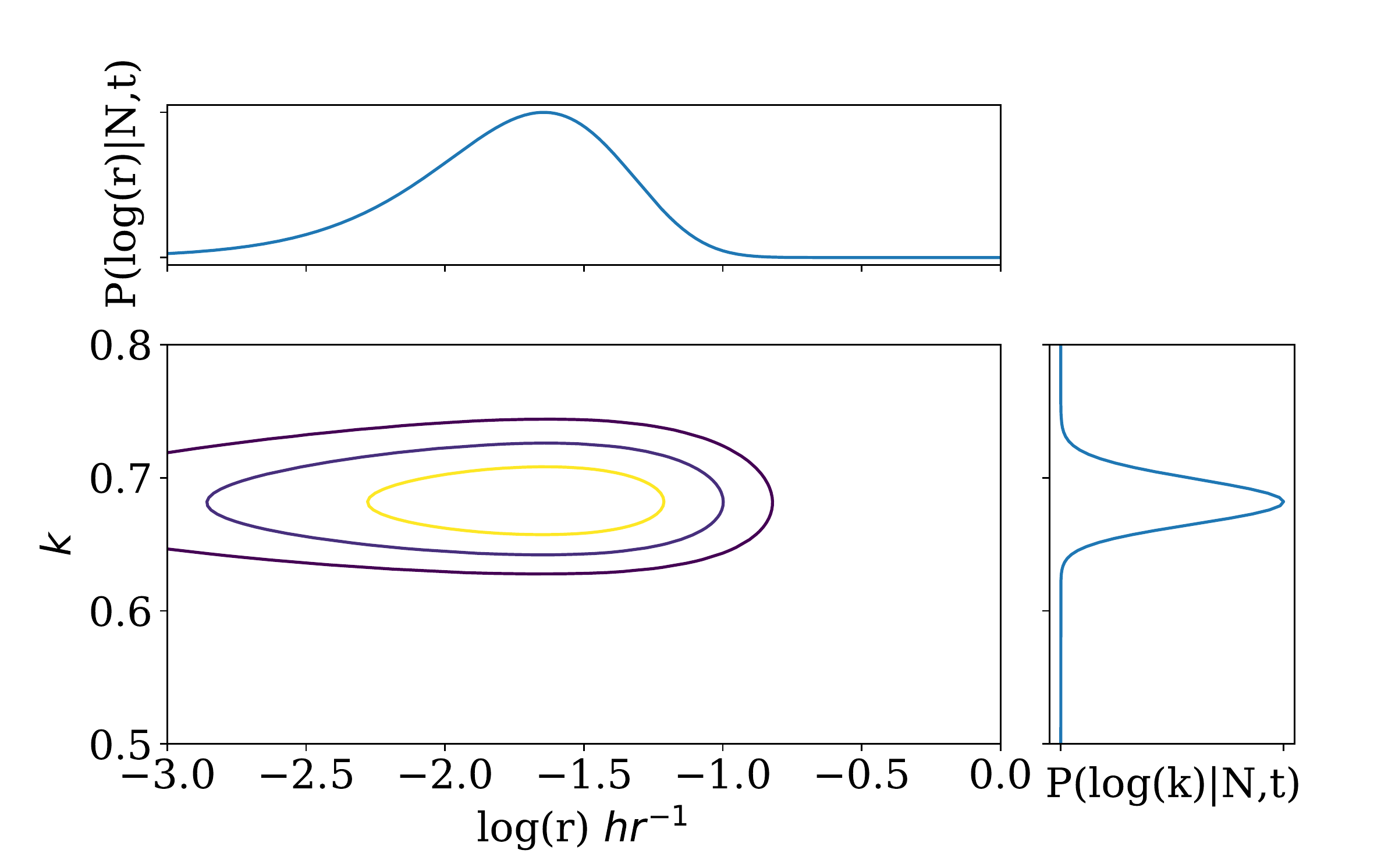}
\includegraphics[width=0.4\textwidth]{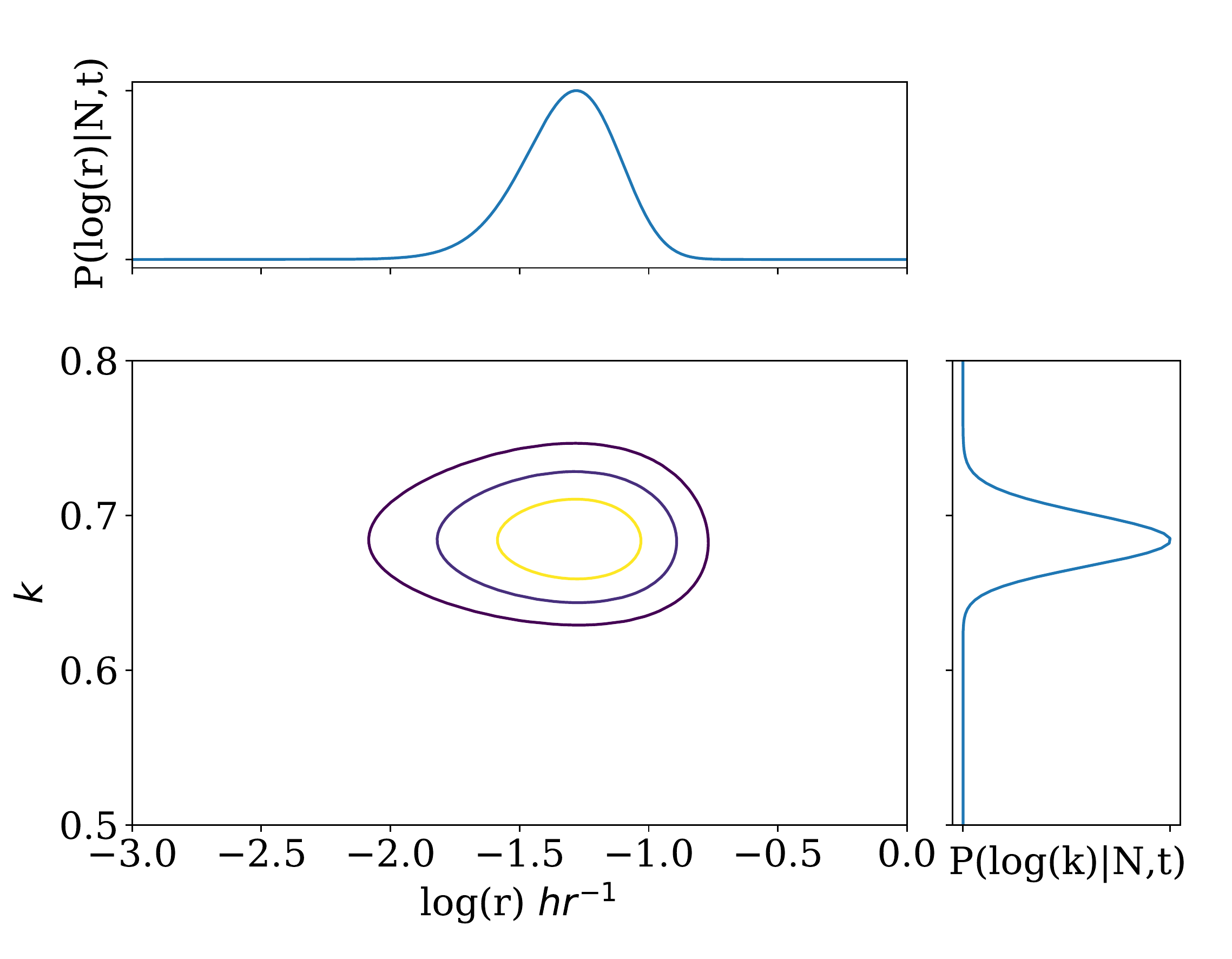}
\caption{Weibull distribution posterior probability distribution for clustering coefficient $k$ and burst rate $r$ for \rseven{} (top panel) and \rsixteen{} (bottom panel). Contours are at 68\%, 95\%, and 99\% probability. The insets are marginalized posteriors for $\log{(r)}$ and $\log{(k)}$. The \cite{oyp18} implementation of the Weibull distribution finds a moderate amount of clustering and burst rates above 1 Jy similar to the Poissonian confidence interval burst rates discussed earlier. The preponderance of non-detections means that there is little distinction between maximum likelihood $k$ and $r$ values for the two sources. \label{fig:weibull}}
\end{figure}

In Table \ref{tab:weibull}, we present the posterior mean values for $r$ and for $k$ as calculated using this method. In Figure \ref{fig:weibull}, we present a contour plot with the 68\%, 95\%, and 99\% contours for $k$ and $r$, for the two known repeating sources. 

\begin{table}
\begin{center}
\caption{Posterior means and uncertainties for Weibull distribution \cfrb{} + AO repetition rates\tablenotemark{a} \label{tab:weibull}}
\begin{tabular}{lcc}
\hline\hline
Source  & $\log{(r)} (hr^{-1})$  & $\log{(k)}$ \\ \hline 
FRB 20190116A \tablenotemark{b}  & $ -1.64_{-0.34}^{+0.26}$ & $-0.166_{-0.008}^{0.008}$  \\ \\
FRB 20190117A  \tablenotemark{b} &  $-1.28_{-0.16}^{+0.15}$ & $-0.164_{-0.010}^{0.008}$  \\ \\
FRB 20190109A  & $-2.08_{-0.45}^{+ 0.33}$ & $-0.164_{-0.010}^{0.008}$  \\ \\
FRB 20181030E  & $ -1.22_{-0.444}^{+0.333}$ & $-0.164_{-0.010}^{0.008}$  \\ \\
FRB 20181125A  & $-1.85_{-0.45}^{+0.33}$ & $-0.164_{-0.010}^{0.008}$  \\ \\
FRB 20181226B  & $ -1.92_{-0.44}^{+0.34} $ & $-0.164_{-0.010}^{0.008}$  \\ \\
FRB 20190124C  & $-1.82^{+0.33}_{-0.44}$ & $-0.164_{-0.010}^{0.008}$  \\ \\
FRB 20190111A  & $-1.82^{+0.33}_{-0.44}$ & $-0.164_{-0.010}^{0.008}$  \\ \\
FRB 20181224E  & $-1.82^{+0.33}_{-0.44}$ & $-0.164_{-0.010}^{0.008}$  \\ \\ \hline
\end{tabular}
\end{center}
\tablenotetext{a}{Above 1 Jy}
\tablenotetext{b}{Known repeater}

\end{table}

Based on the rate analyses, we can say that the repetition rates are generally low (on the order of $10^{-1}$ to $10^{-2} \; \text{hr}^{-1}$). For comparison, analysis of observations of \rone{} with the Five Hundred Metre Aperture Synthesis Telescope (FAST) found found a burst rate of $736^{+26.55}_{-28.90}$ day$^{-1}$ using the same approach in \cite{zww+21}. As in the Poisson analysis, using a scaled exposure time increases the repetition rate for FRB 20181030E due to its narrow width. 

Though there may be a moderate clustering effect, it is immediately obvious that the values determined by the Weibull analysis for clustering coefficient $k$ are nearly identical for all sources, and are identical for all Group B sources. This suggests that the Weibull analysis is not providing additional, physically meaningful description of the burst repetition vs. the simpler Poissonian analysis as it cannot distinguish different clustering behavior among the sources. 

This likely occurs because of the preponderance of non-detections in our dataset. Each source's list of observation includes approximately 760 observations with \cfrb{}. Only one to six of these \cfrb{} observations contain a detection. \cfrb{} non-detections are the dominant case (as opposed to \cfrb{} detections or Arecibo Observatory non-detections). Therefore, mathematically, we are looking at nearly identical datasets, particularly for the seven sources with only one detection and over 700 non-detections each.

We might hope to see a higher degree of clustering from this analysis, suggesting that we are simply missing clusters of observations by observing at the wrong times. However, we record only one multi-burst observation, for \rseven{}. The Weibull implementation used here does not account for the time between individual observations, so only multi-burst observations appear as clusters. This means \cfrb{}, with its short daily observation duration for these sources is unlikely to detect a cluster of detections; the \rseven{} pair of detections is an unusual case. In principle, the presence of this cluster should distinguish the value of $k$ for \rseven{}. However, there is only one such cluster in 766 observations, so each individual observationaffects the parameters only marginally. It is worth noting that the independent events assumption made in \cite{oyp18} is valid: the time between observations is long relative to the observation duration (one sidereal day compared to approximately three minutes). However, maximizing our understanding of \cfrb{} data will require us to move beyond this simplification.

The excessive imbalance between detections and non-detections and the failure to account for gaps between observations emphasizes that the \cite{oyp18} implementation of the Weibull distribution is likely insufficient for analyses that incorporate \cfrb{} data or data from other all-sky FRB monitors. Though this structure has been commonly used in discussing FRB repetition, the time has come for a more sophisticated model. One method that attempts to address this challenge is that of \cite{jof+20a}, but a full implementation is beyond the scope of this work.

The discovery of periodic activity from \rone{} and \rthree{} also suggests that a Weibull distribution may be an inadequate tool for understanding repetition, as it suggests that a simple repetition rate may be the wrong quantity to consider in discussing frequency of repetition. Our knowledge of repeater periodicity would motivate a further extension of this method to account for the spacing between observations. However, we do not attempt such an extension in this work, as we have no concrete evidence of periodicity within our sources.

\subsection{Comparison with other repeaters \label{sect:compare}}
When noting the very low repetition rates suggested by both Poisson and Weibull analyses, the reader might be tempted to conclude that there was something unusual about these sources - perhaps they were particularly close to the flux density limits of our telescope or perhaps they are particularly distant. Of course, limited observation duration is a concern for a CHIME only analysis, but the inclusion of Arecibo Observatory data in this analysis mitigates that factor, providing focused, long-duration observations of the relevant regions and, for Group A sources, substantially increasing overall exposure time. In this section, we present a brief analysis of these sources in the context of \cfrb{} detected repeating FRBs published in \cite{RN,RN2,jcf+19}. For brevity and for maximum relevance, we compare only CHIME detections and do not include detections from other telescopes such as the ASKAP or the Green Bank Telescope. 

\begin{figure}[hbtp] \label{fig:clancyplot}
\centering
 \includegraphics[width= 0.48\textwidth]{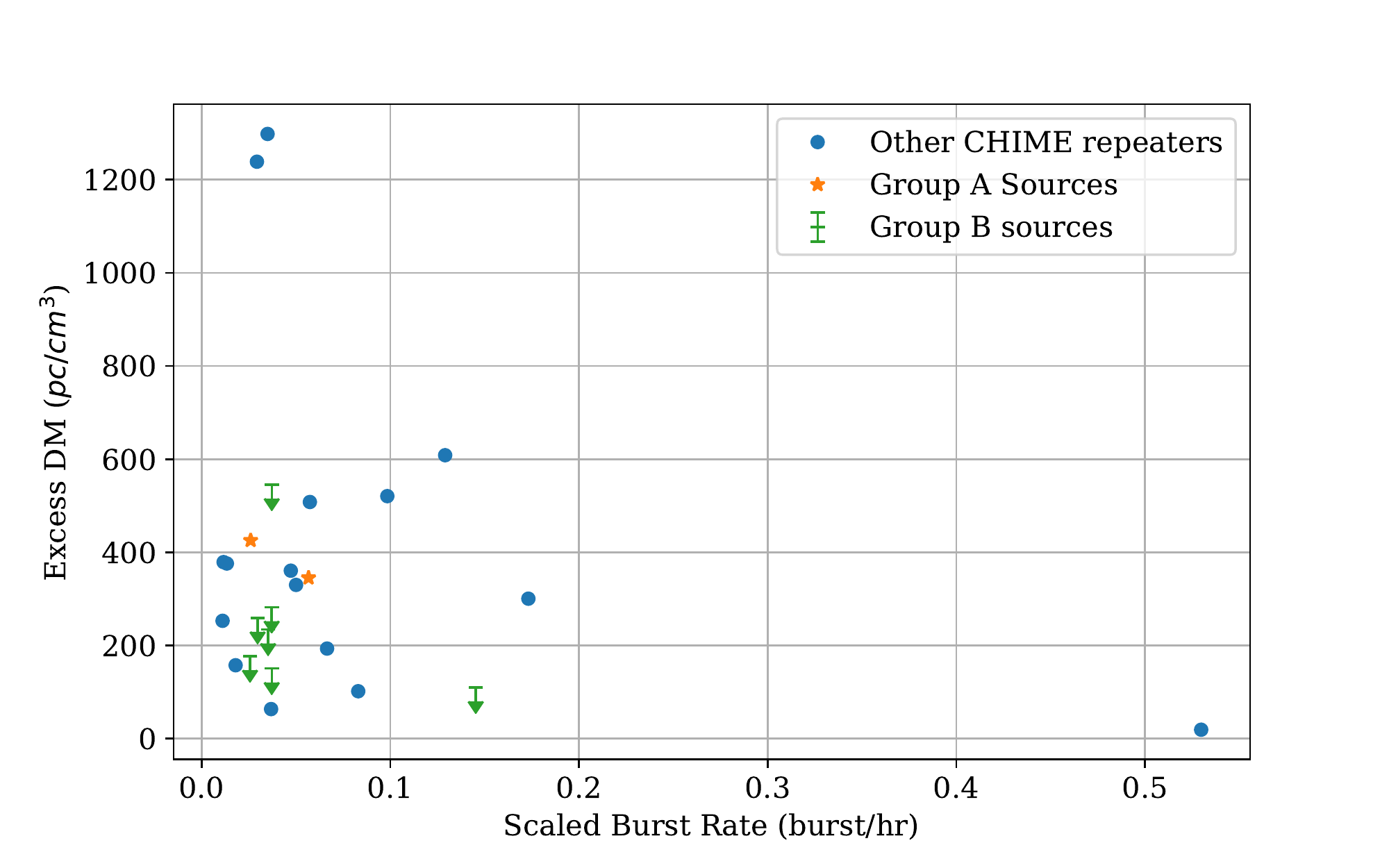} 
 \caption{Excess DM (\dmunits) plotted against burst rate (scaled by sensitivity relative to 1 Jy) for repeaters published in \cite{RN} and \cite{RN2}, as well as \rone{} and Group A and Group B Arecibo Observatory follow-up sources. Group A sources are represented by orange stars, Group B sources by green upper-limit symbols, and other \cfrb{} repeaters by blue dots. Properties of known \cfrb{} repeaters are taken from \cite{cfrbcatalog}. There is no significant correlation between the number of bursts detected by \cfrb{} and excess DM. Additionally, our Arecibo Observatory follow-up sources do not appear to be more distant than observed \cfrb{} repeaters.  \label{fig:clancyplot_dm}}
\end{figure}

In Figure \ref{fig:clancyplot_dm}, we plot excess DM vs. a simple burst rate. As is common in pulsar and FRB astronomy, we are using DM as a proxy for distance. Excess DM is calculated by subtracting the measured Galactic DM from the total measured DM of the FRB. Here, we use the mean of the \cite{ymw17} and \cite{cl02} DM map values for the Galactic DM to these positions. It is important to note that this simple analysis does not account for contributions from the host DM; such a calculation could in principle be conducted with the methods presented in \cite{ckr+22}. The burst rate for our sources is the rate or upper limit from Table \ref{tab:poisson_rep_rate}. The burst rate for other CHIME repeaters is calculated by dividing the number of bursts by the total exposure time August 28, 2018 to May 1, 2021.  

\begin{figure}
\centering
  \includegraphics[width=0.48\textwidth]{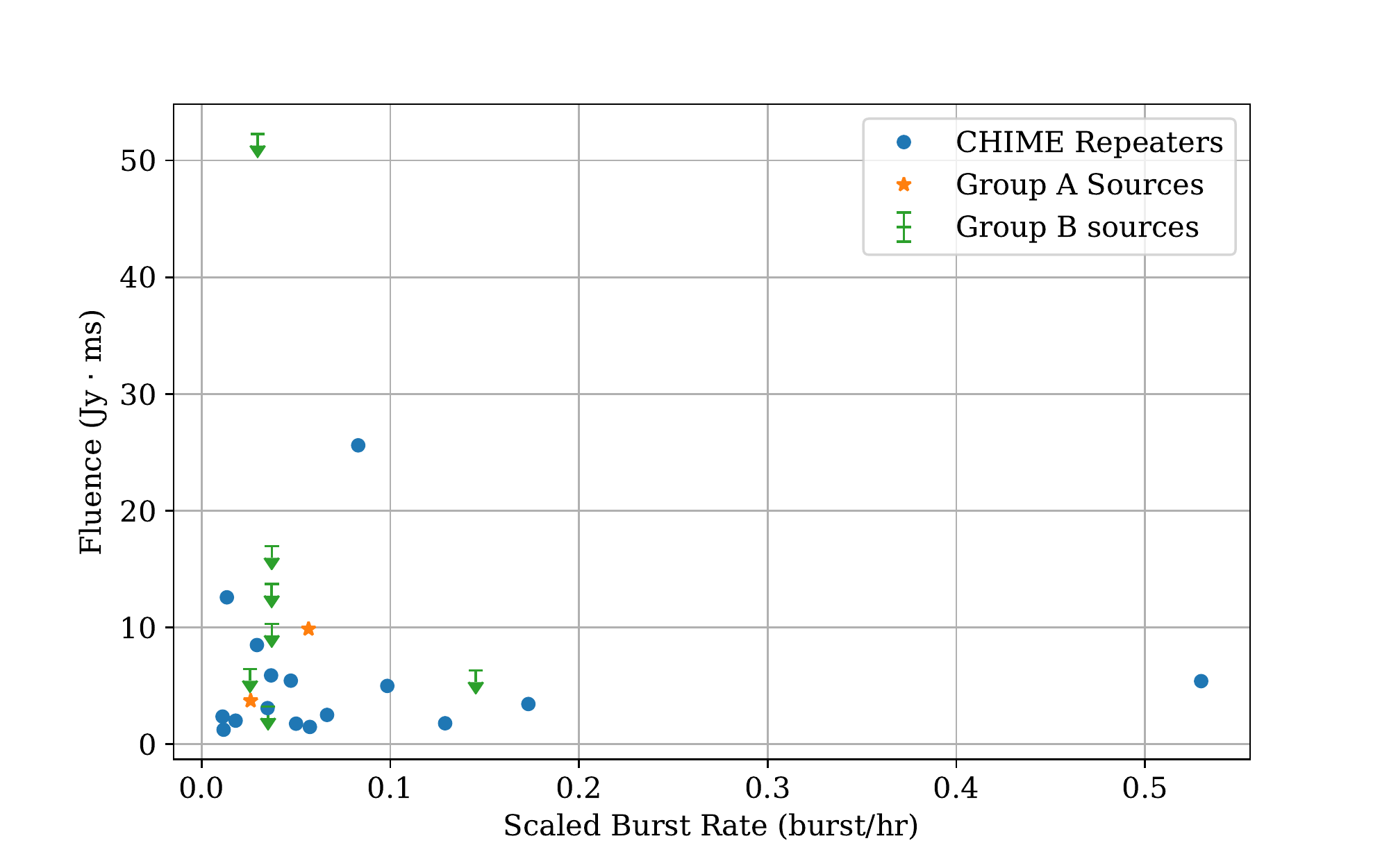}
  \caption{Average \cfrb{} fluence measurement for known \cfrb{} repeaters, Group A sources, and Group B sources, plotted against burst rate scaled above 1 Jy.\label{fig:clancyplot_flux}}
\end{figure}

In Figure \ref{fig:clancyplot_dm}, it is apparent that the burst with the highest number of measured repetitions is also the closest, \rthree{}. However, the relationship between number of repetitions and excess DM is more ambiguous for other bursts. It is certainly the case that our Group A and our Group B sources are not anomalously distant. They appear alongside the bulk of  our published repeaters, suggesting they are not exceptional.

The second question we might ask is are our sources anomalously dim and therefore unlikely to be observed again, i.e., could our detections be the very brightest bursts from this source and thus we are failing to see any further bursts? In Figure \ref{fig:clancyplot_flux}, we plot the burst fluence in Jy ms from \cite{RN} and \cite{RN2} for \cfrb{} repeaters and Group A sources and from \cite{cfrbcatalog} for Group B sources against the scaled burst rates for Group A sources and upper limits for Group B sources. Here, it appears that our sources have slightly higher fluences than \cfrb{} repeaters with similar rates. This suggests that we are not limited to detecting only the brightest bursts from these sources; \cfrb{} has detected repetition from sources with lower fluences. 

Overall, we assert that there is nothing exceptional about these bursts -- they are not particularly distant nor are they particularly dim. This is important as it means we cannot disregard the non-detection of further (or initial) repetition as being simply a function of unusual sources. 

\section{Discussion \& Conclusions \label{sect:discussion}}

This work was conceived with the idea that we could detect further bursts from repeater or repeater-like sources with higher time and frequency resolution and potentially higher sensitivity. We found no such bursts, in 20.1 hours of observation for \rsixteen{}, 27.1 hours of observation for \rseven{}, and 2--5 hours of observation for each Group B sources. However, the non-detection of repeat bursts is instructive and provides an important insight for the study of repeating FRBs broadly: low-burst rate repeating FRBs are a real component of our population, and increased exposure time and improved sensitivity do not ensure the discovery of further repeat bursts.

One potential difficulty in this project was observing cadence. Our observing cadence with the Arecibo Observatory was set essentially at random, based on the idea that we could not predict when FRBs would be active. However, our increasing understanding that FRBs go through active and inactive periods may lead to changes in how we conduct such follow-up observations. Ideally, we would detect this periodicity with a more passive monitor like \cfrb{} and then be able to apply triggered observations as in \cite{cab+20}. But even in the absence of such a program, we may be able to create more efficient programs. It may be better to propose focused blocks of observation, e.g., 1.5 hours per day every day for a week, repeated several times vs. the roughly weekly cadence of these observations.

Another hurdle is our limited knowledge of FRB properties below 400 MHz. Our first sub-400 MHz FRB discoveries were made recently, and we still have only a handful \citep[e.g.][]{pck+20,cab+20,pbp+20,sgp+20, pmb+21}. Of this handful, most are repeat detections of \rthree{}. Though FRBs are detectable at low frequency, it remains to be seen whether lessons from higher frequencies are applicable. 

This work indicates that while morphological studies of  \cfrb{} Catalog 1 have been suggestive of the possibility of two separate populations of FRBs, we should be cautious of over-interpreting  data. Some non-repeating FRBs do appear to have the same morphological characteristics as repeating FRBs. It also suggests that highly clustered and/or very low burst rate repeating FRBs are an important consideration in our population-level understanding of repetition. 

Understanding repeating FRBs across the full range of burst rates has broad implications for our understanding of repeating FRBs at large. The rate distribution has implications for the properties and possible origins of FRBs, with \cite{jof+20b} demonstrating the power of repetition rate to constrain possible progenitor models. \cite{lpw20} proposed that the properties of the first 18 \cfrb{} repeating FRBs could be suggestive of either a shallow energy distribution or a broad distribution of repetition rate; the existence of both low burst rate repeaters like our Group A sources and high burst rate sources like \rone{} and \rthree{} suggests the latter hypothesis.  Further, \cite{james19} suggests that all FRBs cannot be repeating FRBs with rates comparable to \rone{}. This could simply be evidence of multiple populations of repeating and non-repeating FRBs, but would also be consistent with a range of repetition rates amongst repeating FRBs.

The continued non-detection of repetition from morphologically interesting sources is itself very intriguing, but hard to quantify. This may be evidence that all FRBs repeat, via a linkage between repeater like properties and bursts which appear to be one-off, or it may be an indication that we have caught only the tail of brightest bursts (and were unlucky with follow-up observations). 

The demise of Arecibo Observatory means that we cannot directly continue this project. However similar projects could be conducted in the future, using the Five Hundred Metre Aperture Synthesis Telescope (FAST). 

With the release of \cfrb{} Catalog 1 and accompanying meta-analyses, we are beginning to draw educated conclusions about the nature of the FRB population, including the properties of the repeater population. Though we still have much to learn and should draw conclusions cautiously, we also eagerly look forward to further studies of individual repeaters in depth and continued expansion of the known list of repeating FRBs by surveys such as those at FAST, ASKAP, and \cfrb{}.

\begin{acknowledgements}

We acknowledge that CHIME is located on the traditional, ancestral, and unceded territory of the Syilx/Okanagan people. We are grateful to the staff of the Dominion Radio Astrophysical Observatory, which is operated by the National Research Council of Canada.  CHIME is funded by a grant from the Canada Foundation for Innovation (CFI) 2012 Leading Edge Fund (Project 31170) and by contributions from the provinces of British Columbia, Qu\'{e}bec and Ontario. The CHIME/FRB Project is funded by a grant from the CFI 2015 Innovation Fund (Project 33213) and by contributions from the provinces of British Columbia and Qu\'{e}bec, and by the Dunlap Institute for Astronomy and Astrophysics at the University of Toronto. Additional support was provided by the Canadian Institute for Advanced Research (CIFAR), McGill University and the McGill Space Institute, the Trottier Family Foundation, and the University of British Columbia. 

V.M.K. holds the Lorne Trottier Chair in Astrophysics \& Cosmology, a Distinguished James McGill Professorship, and receives support from an NSERC Discovery grant (RGPIN 228738-13), from an R. Howard Webster Foundation Fellowship from CIFAR, and from the FRQNT CRAQ. FRB research at UBC is supported by an NSERC Discovery Grant and by the Canadian Institute for Advanced Research. P.S. and Z.P. are Dunlap Fellows. The Flatiron Institute is supported by the Simons Foundation.
\end{acknowledgements}

\software{\presto{} \cite{presto}, Astropy \cite{astropy}, SciPy \cite{scipy}}

\bibliography{biblio}

\begin{thebibliography}{}
\expandafter\ifx\csname natexlab\endcsname\relax\def\natexlab#1{#1}\fi
\providecommand{\url}[1]{\href{#1}{#1}}
\providecommand{\dodoi}[1]{doi:~\href{http://doi.org/#1}{\nolinkurl{#1}}}
\providecommand{\doeprint}[1]{\href{http://ascl.net/#1}{\nolinkurl{http://ascl.net/#1}}}
\providecommand{\doarXiv}[1]{\href{https://arxiv.org/abs/#1}{\nolinkurl{https://arxiv.org/abs/#1}}}

\bibitem[{{Astropy Collaboration} {et~al.}(2018){Astropy Collaboration},
  {Price-Whelan}, {Sip{\H{o}}cz}, {G{\"u}nther}, {Lim}, {Crawford}, {Conseil},
  {Shupe}, {Craig}, {Dencheva}, {Ginsburg}, {VanderPlas}, {Bradley},
  {P{\'e}rez-Su{\'a}rez}, {de Val-Borro}, {Aldcroft}, {Cruz}, {Robitaille},
  {Tollerud}, {Ardelean}, {Babej}, {Bach}, {Bachetti}, {Bakanov}, {Bamford},
  {Barentsen}, {Barmby}, {Baumbach}, {Berry}, {Biscani}, {Boquien}, {Bostroem},
  {Bouma}, {Brammer}, {Bray}, {Breytenbach}, {Buddelmeijer}, {Burke},
  {Calderone}, {Cano Rodr{\'\i}guez}, {Cara}, {Cardoso}, {Cheedella}, {Copin},
  {Corrales}, {Crichton}, {D'Avella}, {Deil}, {Depagne}, {Dietrich}, {Donath},
  {Droettboom}, {Earl}, {Erben}, {Fabbro}, {Ferreira}, {Finethy}, {Fox},
  {Garrison}, {Gibbons}, {Goldstein}, {Gommers}, {Greco}, {Greenfield},
  {Groener}, {Grollier}, {Hagen}, {Hirst}, {Homeier}, {Horton}, {Hosseinzadeh},
  {Hu}, {Hunkeler}, {Ivezi{\'c}}, {Jain}, {Jenness}, {Kanarek}, {Kendrew},
  {Kern}, {Kerzendorf}, {Khvalko}, {King}, {Kirkby}, {Kulkarni}, {Kumar},
  {Lee}, {Lenz}, {Littlefair}, {Ma}, {Macleod}, {Mastropietro}, {McCully},
  {Montagnac}, {Morris}, {Mueller}, {Mumford}, {Muna}, {Murphy}, {Nelson},
  {Nguyen}, {Ninan}, {N{\"o}the}, {Ogaz}, {Oh}, {Parejko}, {Parley}, {Pascual},
  {Patil}, {Patil}, {Plunkett}, {Prochaska}, {Rastogi}, {Reddy Janga},
  {Sabater}, {Sakurikar}, {Seifert}, {Sherbert}, {Sherwood-Taylor}, {Shih},
  {Sick}, {Silbiger}, {Singanamalla}, {Singer}, {Sladen}, {Sooley},
  {Sornarajah}, {Streicher}, {Teuben}, {Thomas}, {Tremblay}, {Turner},
  {Terr{\'o}n}, {van Kerkwijk}, {de la Vega}, {Watkins}, {Weaver}, {Whitmore},
  {Woillez}, {Zabalza}, \& {Astropy Contributors}}]{astropy}
{Astropy Collaboration}, {Price-Whelan}, A.~M., {Sip{\H{o}}cz}, B.~M., {et~al.}
  2018, \aj, 156, 123, \dodoi{10.3847/1538-3881/aabc4f}

\bibitem[{{Caleb} {et~al.}(2019{\natexlab{a}}){Caleb}, {Stappers}, {Rajwade},
  \& Flynn}]{csr19}
{Caleb}, M., {Stappers}, B., {Rajwade}, K., \& Flynn, C. 2019{\natexlab{a}},
  \mnras, 484, 5500

\bibitem[{{Caleb} {et~al.}(2019{\natexlab{b}}){Caleb}, {Stappers}, {Rajwade},
  \& {Flynn}}]{csr+19}
{Caleb}, M., {Stappers}, B.~W., {Rajwade}, K., \& {Flynn}, C.
  2019{\natexlab{b}}, \mnras, 484, 5500, \dodoi{10.1093/mnras/stz386}

\bibitem[{{Chawla} {et~al.}(2020){Chawla}, {Andersen}, {Bhardwaj}, {Fonseca},
  {Josephy}, {Kaspi}, {Michilli}, {Pleunis}, {Bandura}, {Bassa}, {Boyle},
  {Brar}, {Cassanelli}, {Cubranic}, {Dobbs}, {Dong}, {Gaensler}, {Good},
  {Hessels}, {Landecker}, {Leung}, {Li}, {Lin}, {Masui}, {Mckinven},
  {Mena-Parra}, {Merryfield}, {Meyers}, {Naidu}, {Ng}, {Patel},
  {Rafiei-Ravandi}, {Rahman}, {Sanghavi}, {Scholz}, {Shin}, {Smith}, {Stairs},
  {Tendulkar}, \& {Vanderlinde}}]{cab+20}
{Chawla}, P., {Andersen}, B.~C., {Bhardwaj}, M., {et~al.} 2020, \apjl, 896,
  L41, \dodoi{10.3847/2041-8213/ab96bf}

\bibitem[{{Chawla} {et~al.}(2022){Chawla}, {Kaspi}, {Ransom}, {Bhardwaj},
  {Boyle}, {Breitman}, {Cassanelli}, {Cubranic}, {Dong}, {Fonseca}, {Gaensler},
  {Giri}, {Josephy}, {Kaczmarek}, {Leung}, {Masui}, {Mena-Parra}, {Merryfield},
  {Michilli}, {M{\"u}nchmeyer}, {Ng}, {Patel}, {Pearlman}, {Petroff},
  {Pleunis}, {Rahman}, {Sanghavi}, {Shin}, {Smith}, {Stairs}, \&
  {Tendulkar}}]{ckr+22}
{Chawla}, P., {Kaspi}, V.~M., {Ransom}, S.~M., {et~al.} 2022, \apj, 927, 35,
  \dodoi{10.3847/1538-4357/ac49e1}

\bibitem[{{CHIME/FRB Collaboration}(2018)}]{chimefrb}
{CHIME/FRB Collaboration}. 2018, \apj, 863, 48,
  \dodoi{10.3847/1538-4357/aad188}

\bibitem[{{CHIME/FRB Collaboration}(2019{\natexlab{a}})}]{RN}
---. 2019{\natexlab{a}}, \apjl, 885, L24, \dodoi{10.3847/2041-8213/ab4a80}

\bibitem[{{CHIME/FRB Collaboration}(2019{\natexlab{b}})}]{R2}
---. 2019{\natexlab{b}}, \nat, 566, 235, \dodoi{10.1038/s41586-018-0864-x}

\bibitem[{{CHIME/FRB Collaboration}(2020)}]{R3periodicity}
---. 2020, \nat, 582, 351, \dodoi{10.1038/s41586-020-2398-2}

\bibitem[{{CHIME/FRB Collaboration}(2021)}]{cfrbcatalog}
---. 2021, \apjs, 257, 59, \dodoi{10.3847/1538-4365/ac33ab}

\bibitem[{{CHIME/FRB Collaboration} {et~al.}(2021){CHIME/FRB Collaboration},
  {Andersen}, {Bandura}, {Bhardwaj}, {Boyle}, {Brar}, {Breitman}, {Cassanelli},
  {Chatterjee}, {Chawla}, {Cliche}, {Cubranic}, {Curtin}, {Deng}, {Dobbs},
  {Dong}, {Fonseca}, {Gaensler}, {Giri}, {Good}, {Hill}, {Josephy},
  {Kaczmarek}, {Kader}, {Kania}, {Kaspi}, {Leung}, {Li}, {Lin}, {Masui},
  {Mckinven}, {Mena-Parra}, {Merryfield}, {Meyers}, {Michilli}, {Naidu},
  {Newburgh}, {Ng}, {Ordog}, {Patel}, {Pearlman}, {Pen}, {Petroff}, {Pleunis},
  {Rafiei-Ravandi}, {Rahman}, {Ransom}, {Renard}, {Sanghavi}, {Scholz}, {Shaw},
  {Shin}, {Siegel}, {Singh}, {Smith}, {Stairs}, {Tan}, {Tendulkar},
  {Vanderlinde}, {Wiebe}, {Wulf}, \& {Zwaniga}}]{millisecondperiodicity}
{CHIME/FRB Collaboration}, {Andersen}, B.~C., {Bandura}, K., {et~al.} 2021,
  arXiv e-prints, arXiv:2107.08463.
\newblock \doarXiv{2107.08463}

\bibitem[{{Connor} {et~al.}(2020){Connor}, {Miller}, \& {Gardenier}}]{cmg20}
{Connor}, L., {Miller}, M.~C., \& {Gardenier}, D.~W. 2020, \mnras, 497, 3076,
  \dodoi{10.1093/mnras/staa2074}

\bibitem[{{Connor} \& {Petroff}(2018)}]{cp18}
{Connor}, L., \& {Petroff}, E. 2018, \apjl, 861, L1,
  \dodoi{10.3847/2041-8213/aacd02}

\bibitem[{{Cordes} \& {Lazio}(2002)}]{cl02}
{Cordes}, J.~M., \& {Lazio}, T.~J.~W. 2002, arXiv e-prints, astro.
\newblock \doarXiv{astro-ph/0207156}

\bibitem[{{Cordes} \& {McLaughlin}(2003)}]{cm+03}
{Cordes}, J.~M., \& {McLaughlin}, M.~A. 2003, \apj, 596, 1142,
  \dodoi{10.1086/378231}

\bibitem[{{Cruces} {et~al.}(2021){Cruces}, {Spitler}, {Scholz}, {Lynch},
  {Seymour}, {Hessels}, {Gouiff{\'e}s}, {Hilmarsson}, {Kramer}, \&
  {Munjal}}]{css+20}
{Cruces}, M., {Spitler}, L.~G., {Scholz}, P., {et~al.} 2021, \mnras, 500, 448,
  \dodoi{10.1093/mnras/staa3223}

\bibitem[{{Fonseca} {et~al.}(2020){Fonseca}, {Andersen}, {Bhardwaj}, {Chawla},
  {Good}, {Josephy}, {Kaspi}, {Masui}, {Mckinven}, {Michilli}, {Pleunis},
  {Shin}, {Tendulkar}, {Bandura}, {Boyle}, {Brar}, {Cassanelli}, {Cubranic},
  {Dobbs}, {Dong}, {Gaensler}, {Hinshaw}, {Landecker}, {Leung}, {Li}, {Lin},
  {Mena-Parra}, {Merryfield}, {Naidu}, {Ng}, {Patel}, {Pen}, {Rafiei-Ravandi},
  {Rahman}, {Ransom}, {Scholz}, {Smith}, {Stairs}, {Vanderlinde}, {Yadav}, \&
  {Zwaniga}}]{RN2}
{Fonseca}, E., {Andersen}, B.~C., {Bhardwaj}, M., {et~al.} 2020, \apjl, 891,
  L6, \dodoi{10.3847/2041-8213/ab7208}

\bibitem[{{Hessels} {et~al.}(2019){Hessels}, {Spitler}, {Seymour}, {Cordes},
  {Michilli}, {Lynch}, {Gourdji}, {Archibald}, {Bassa}, {Bower}, {Chatterjee},
  {Connor}, {Crawford}, {Deneva}, {Gajjar}, {Kaspi}, {Keimpema}, {Law},
  {Marcote}, {McLaughlin}, {Paragi}, {Petroff}, {Ransom}, {Scholz}, {Stappers},
  \& {Tendulkar}}]{hss+19}
{Hessels}, J.~W.~T., {Spitler}, L.~G., {Seymour}, A.~D., {et~al.} 2019, \apjl,
  876, L23, \dodoi{10.3847/2041-8213/ab13ae}

\bibitem[{{Hilmarsson} {et~al.}(2021){Hilmarsson}, {Michilli}, {Spitler},
  {Wharton}, {Demorest}, {Desvignes}, {Gourdji}, {Hackstein}, {Hessels},
  {Nimmo}, {Seymour}, {Kramer}, \& {Mckinven}}]{hmsw+21}
{Hilmarsson}, G.~H., {Michilli}, D., {Spitler}, L.~G., {et~al.} 2021, \apjl,
  908, L10, \dodoi{10.3847/2041-8213/abdec0}

\bibitem[{{James}(2019)}]{james19}
{James}, C.~W. 2019, \mnras, 486, 5934, \dodoi{10.1093/mnras/stz1224}

\bibitem[{{James} {et~al.}(2020{\natexlab{a}}){James}, {Os{\l}owski}, {Flynn},
  {Kumar}, {Bannister}, {Bhandari}, {Farah}, {Kerr}, {Lorimer}, {Macquart},
  {Ng}, {Phillips}, {Price}, {Qiu}, {Shannon}, \& {Spiewak}}]{jof+20a}
{James}, C.~W., {Os{\l}owski}, S., {Flynn}, C., {et~al.} 2020{\natexlab{a}},
  \mnras, 495, 2416, \dodoi{10.1093/mnras/staa1361}

\bibitem[{{James} {et~al.}(2020{\natexlab{b}}){James}, {Os{\l}owski}, {Flynn},
  {Kumar}, {Bannister}, {Bhandari}, {Farah}, {Kerr}, {Lorimer}, {Macquart},
  {Ng}, {Phillips}, {Price}, {Qiu}, {Shannon}, \& {Spiewak}}]{jof+20b}
---. 2020{\natexlab{b}}, \apjl, 895, L22, \dodoi{10.3847/2041-8213/ab8f99}

\bibitem[{{Josephy} {et~al.}(2019){Josephy}, {Chawla}, {Fonseca}, {Ng},
  {Patel}, {Pleunis}, {Scholz}, {Andersen}, {Bandura}, {Bhardwaj}, {Boyce},
  {Boyle}, {Brar}, {Cubranic}, {Dobbs}, {Gaensler}, {Gill}, {Giri}, {Good},
  {Halpern}, {Hinshaw}, {Kaspi}, {Landecker}, {Lang}, {Lin}, {Masui},
  {Mckinven}, {Mena-Parra}, {Merryfield}, {Michilli}, {Milutinovic}, {Naidu},
  {Pen}, {Rafiei-Ravandi}, {Rahman}, {Ransom}, {Renard}, {Siegel}, {Smith},
  {Stairs}, {Tendulkar}, {Vanderlinde}, {Yadav}, \& {Zwaniga}}]{jcf+19}
{Josephy}, A., {Chawla}, P., {Fonseca}, E., {et~al.} 2019, \apjl, 882, L18,
  \dodoi{10.3847/2041-8213/ab2c00}

\bibitem[{{Kraft} {et~al.}(1991){Kraft}, {Burrows}, \& {Nousek}}]{kbn91}
{Kraft}, R.~P., {Burrows}, D.~N., \& {Nousek}, J.~A. 1991, \apj, 374, 344,
  \dodoi{10.1086/170124}

\bibitem[{Kumar {et~al.}(2019)Kumar, Shannon, Os{\l}owski, {et~al.}}]{kso+19}
Kumar, P., Shannon, R., Os{\l}owski, S., {et~al.} 2019, The Astrophysical
  Journal Letters, 887, L30

\bibitem[{{Lanman} {et~al.}(2021){Lanman}, {Andersen}, {Chawla}, {Josephy},
  {Noble}, {Kaspi}, {Bandura}, {Bhardwaj}, {Boyle}, {Brar}, {Breitman},
  {Cassanelli}, {Dong}, {Fonseca}, {Gaensler}, {Good}, {Kaczmarek}, {Leung},
  {Masui}, {Meyers}, {Ng}, {Patel}, {Pearlman}, {Petroff}, {Pleunis},
  {Rafiei-Ravandi}, {Rahman}, {Sanghavi}, {Scholz}, {Shin}, {Stairs},
  {Tendulkar}, \& {Zwaniga}}]{r67}
{Lanman}, A.~E., {Andersen}, B.~C., {Chawla}, P., {et~al.} 2021, arXiv
  e-prints, arXiv:2109.09254.
\newblock \doarXiv{2109.09254}

\bibitem[{{Li} {et~al.}(2021){Li}, {Wang}, {Zhu}, {Zhang}, {Zhang}, {Duan},
  {Zhang}, {Feng}, {Tang}, {Chatterjee}, {Cordes}, {Cruces}, {Dai}, {Gajjar},
  {Hobbs}, {Jin}, {Kramer}, {Lorimer}, {Miao}, {Niu}, {Niu}, {Pan}, {Qian},
  {Spitler}, {Werthimer}, {Zhang}, {Wang}, {Xie}, {Yue}, {Zhang}, {Zhi}, \&
  {Zhu}}]{lwz+21}
{Li}, D., {Wang}, P., {Zhu}, W.~W., {et~al.} 2021, \nat, 598, 267,
  \dodoi{10.1038/s41586-021-03878-5}

\bibitem[{{Lorimer} {et~al.}(2007){Lorimer}, {Bailes}, {McLaughlin},
  {Narkevic}, \& {Crawford}}]{lbm+07}
{Lorimer}, D.~R., {Bailes}, M., {McLaughlin}, M.~A., {Narkevic}, D.~J., \&
  {Crawford}, F. 2007, Science, 318, 777, \dodoi{10.1126/science.1147532}

\bibitem[{{Lorimer} \& {Kramer}(2005)}]{lk04}
{Lorimer}, D.~R., \& {Kramer}, M. 2005, {Handbook of Pulsar Astronomy}, Vol.~4

\bibitem[{{Lu} {et~al.}(2020){Lu}, {Piro}, \& {Waxman}}]{lpw20}
{Lu}, W., {Piro}, A.~L., \& {Waxman}, E. 2020, \mnras, 498, 1973,
  \dodoi{10.1093/mnras/staa2397}

\bibitem[{{Majid} {et~al.}(2021){Majid}, {Pearlman}, {Prince}, {Wharton},
  {Naudet}, {Bansal}, {Connor}, {Bhardwaj}, \& {Tendulkar}}]{mpp+21}
{Majid}, W.~A., {Pearlman}, A.~B., {Prince}, T.~A., {et~al.} 2021, \apjl, 919,
  L6, \dodoi{10.3847/2041-8213/ac1921}

\bibitem[{{Michilli} {et~al.}(2018){Michilli}, {Seymour}, {Hessels}, {Spitler},
  {Gajjar}, {Archibald}, {Bower}, {Chatterjee}, {Cordes}, {Gourdji}, {Heald},
  {Kaspi}, {Law}, {Sobey}, {Adams}, {Bassa}, {Bogdanov}, {Brinkman},
  {Demorest}, {Fernandez}, {Hellbourg}, {Lazio}, {Lynch}, {Maddox}, {Marcote},
  {McLaughlin}, {Paragi}, {Ransom}, {Scholz}, {Siemion}, {Tendulkar}, {van
  Rooy}, {Wharton}, \& {Whitlow}}]{msh+18}
{Michilli}, D., {Seymour}, A., {Hessels}, J.~W.~T., {et~al.} 2018, \nat, 553,
  182, \dodoi{10.1038/nature25149}

\bibitem[{{Michilli} {et~al.}(2021){Michilli}, {Masui}, {Mckinven}, {Cubranic},
  {Bruneault}, {Brar}, {Patel}, {Boyle}, {Stairs}, {Renard}, {Bandura},
  {Berger}, {Breitman}, {Cassanelli}, {Dobbs}, {Kaspi}, {Leung}, {Mena-Parra},
  {Pleunis}, {Russell}, {Scholz}, {Siegel}, {Tendulkar}, \&
  {Vanderlinde}}]{mmm+21}
{Michilli}, D., {Masui}, K.~W., {Mckinven}, R., {et~al.} 2021, \apj, 910, 147,
  \dodoi{10.3847/1538-4357/abe626}

\bibitem[{{Nimmo} {et~al.}(2021){Nimmo}, {Hessels}, {Keimpema}, {Archibald},
  {Cordes}, {Karuppusamy}, {Kirsten}, {Li}, {Marcote}, \& {Paragi}}]{nhk+21}
{Nimmo}, K., {Hessels}, J.~W.~T., {Keimpema}, A., {et~al.} 2021, Nature
  Astronomy, 5, 594, \dodoi{10.1038/s41550-021-01321-3}

\bibitem[{{Oppermann} {et~al.}(2018){Oppermann}, {Yu}, \& {Pen}}]{oyp18}
{Oppermann}, N., {Yu}, H.-R., \& {Pen}, U.-L. 2018, \mnras, 475, 5109,
  \dodoi{10.1093/mnras/sty004}

\bibitem[{{Parent} {et~al.}(2020){Parent}, {Chawla}, {Kaspi}, {Agazie},
  {Blumer}, {DeCesar}, {Fiore}, {Fonseca}, {Hessels}, {Kaplan}, {Kondratiev},
  {LaRose}, {Levin}, {Lewis}, {Lynch}, {McEwen}, {McLaughlin}, {Mingyar}, {Al
  Noori}, {Ransom}, {Roberts}, {Schmiedekamp}, {Schmiedekamp}, {Siemens},
  {Spiewak}, {Stairs}, {Surnis}, {Swiggum}, \& {van Leeuwen}}]{pck+20}
{Parent}, E., {Chawla}, P., {Kaspi}, V.~M., {et~al.} 2020, \apj, 904, 92,
  \dodoi{10.3847/1538-4357/abbdf6}

\bibitem[{{Pilia} {et~al.}(2020){Pilia}, {Burgay}, {Possenti}, {Ridolfi},
  {Gajjar}, {Corongiu}, {Perrodin}, {Bernardi}, {Naldi}, {Pupillo},
  {Ambrosino}, {Bianchi}, {Burtovoi}, {Casella}, {Casentini}, {Cecconi},
  {Ferrigno}, {Fiori}, {Gendreau}, {Ghedina}, {Naletto}, {Nicastro}, {Ochner},
  {Palazzi}, {Panessa}, {Papitto}, {Pittori}, {Rea}, {Castillo}, {Savchenko},
  {Setti}, {Tavani}, {Trois}, {Trudu}, {Turatto}, {Ursi}, {Verrecchia}, \&
  {Zampieri}}]{pbp+20}
{Pilia}, M., {Burgay}, M., {Possenti}, A., {et~al.} 2020, \apjl, 896, L40,
  \dodoi{10.3847/2041-8213/ab96c0}

\bibitem[{{Pleunis} {et~al.}(2021{\natexlab{a}}){Pleunis}, {Good}, {Kaspi},
  {Mckinven}, {Ransom}, {Scholz}, {Bandura}, {Bhardwaj}, {Boyle}, {Brar},
  {Cassanelli}, {Chawla}, {(Adam) Dong}, {Fonseca}, {Gaensler}, {Josephy},
  {Kaczmarek}, {Leung}, {Lin}, {Masui}, {Mena-Parra}, {Michilli}, {Ng},
  {Patel}, {Rafiei-Ravandi}, {Rahman}, {Sanghavi}, {Shin}, {Smith}, {Stairs},
  \& {Tendulkar}}]{pgk+21}
{Pleunis}, Z., {Good}, D.~C., {Kaspi}, V.~M., {et~al.} 2021{\natexlab{a}},
  \apj, 923, 1, \dodoi{10.3847/1538-4357/ac33ac}

\bibitem[{{Pleunis} {et~al.}(2021{\natexlab{b}}){Pleunis}, {Michilli}, {Bassa},
  {Hessels}, {Naidu}, {Andersen}, {Chawla}, {Fonseca}, {Gopinath}, {Kaspi},
  {Kondratiev}, {Li}, {Bhardwaj}, {Boyle}, {Brar}, {Cassanelli}, {Gupta},
  {Josephy}, {Karuppusamy}, {Keimpema}, {Kirsten}, {Leung}, {Marcote}, {Masui},
  {Mckinven}, {Meyers}, {Ng}, {Nimmo}, {Paragi}, {Rahman}, {Scholz}, {Shin},
  {Smith}, {Stairs}, \& {Tendulkar}}]{pmb+21}
{Pleunis}, Z., {Michilli}, D., {Bassa}, C.~G., {et~al.} 2021{\natexlab{b}},
  \apjl, 911, L3, \dodoi{10.3847/2041-8213/abec72}

\bibitem[{{Rajwade} {et~al.}(2020){Rajwade}, {Mickaliger}, {Stappers},
  {Morello}, {Agarwal}, {Bassa}, {Breton}, {Caleb}, {Karastergiou}, {Keane}, \&
  {Lorimer}}]{rms+20}
{Rajwade}, K.~M., {Mickaliger}, M.~B., {Stappers}, B.~W., {et~al.} 2020,
  \mnras, 495, 3551, \dodoi{10.1093/mnras/staa1237}

\bibitem[{{Ransom}(2011)}]{presto}
{Ransom}, S. 2011, {PRESTO: PulsaR Exploration and Search TOolkit}.
\newblock \doeprint{1107.017}

\bibitem[{{Ravi}(2019)}]{ravi19}
{Ravi}, V. 2019, \apj, 872, 88

\bibitem[{{Rickett}(1969)}]{rickett69}
{Rickett}, B.~J. 1969, \nat, 221, 158, \dodoi{10.1038/221158a0}

\bibitem[{{Sand} {et~al.}(2020){Sand}, {Gajjar}, {Pilia}, {Kudale}, {Joshi},
  {Jagtap}, {Ray}, {Deshpande}, {Bijay}, {Dey}, {Kalita}, {Bandyopadhyay},
  {Jena}, {Bhattacharya}, {Waratkar}, {Wagle}, {Singha}, {Bagchi}, {Surnis},
  {Bhat}, {Mishra}, {Konar}, \& {Maan}}]{sgp+20}
{Sand}, K.~R., {Gajjar}, V., {Pilia}, M., {et~al.} 2020, The Astronomer's
  Telegram, 13781, 1

\bibitem[{{Scholz} {et~al.}(2016){Scholz}, {Spitler}, {Hessels}, {Chatterjee},
  {Cordes}, {Kaspi}, {Wharton}, {Bassa}, {Bogdanov}, {Camilo}, {Crawford},
  {Deneva}, {van Leeuwen}, {Lynch}, {Madsen}, {McLaughlin}, {Mickaliger},
  {Parent}, {Patel}, {Ransom}, {Seymour}, {Stairs}, {Stappers}, \&
  {Tendulkar}}]{ssh+16b}
{Scholz}, P., {Spitler}, L.~G., {Hessels}, J.~W.~T., {et~al.} 2016, \apj, 833,
  177, \dodoi{10.3847/1538-4357/833/2/177}

\bibitem[{{Spitler} {et~al.}(2016){Spitler}, {Scholz}, {Hessels}, {Bogdanov},
  {Brazier}, {Camilo}, {Chatterjee}, {Cordes}, {Crawford}, {Deneva}, {Ferdman},
  {Freire}, {Kaspi}, {Lazarus}, {Lynch}, {Madsen}, {McLaughlin}, {Patel},
  {Ransom}, {Seymour}, {Stairs}, {Stappers}, {van Leeuwen}, \& {Zhu}}]{ssh+16a}
{Spitler}, L.~G., {Scholz}, P., {Hessels}, J.~W.~T., {et~al.} 2016, Nature,
  531, 202, \dodoi{10.1038/nature17168}

\bibitem[{{Virtanen} {et~al.}(2020){Virtanen}, {Gommers}, {Oliphant},
  {Haberland}, {Reddy}, {Cournapeau}, {Burovski}, {Peterson}, {Weckesser},
  {Bright}, {van der Walt}, {Brett}, {Wilson}, {Millman}, {Mayorov}, {Nelson},
  {Jones}, {Kern}, {Larson}, {Carey}, {Polat}, {Feng}, {Moore}, {VanderPlas},
  {Laxalde}, {Perktold}, {Cimrman}, {Henriksen}, {Quintero}, {Harris},
  {Archibald}, {Ribeiro}, {Pedregosa}, {van Mulbregt}, \& {SciPy 1. 0
  Contributors}}]{scipy}
{Virtanen}, P., {Gommers}, R., {Oliphant}, T.~E., {et~al.} 2020, Nature
  Methods, 17, 261, \dodoi{10.1038/s41592-019-0686-2}

\bibitem[{{Yao} {et~al.}(2017){Yao}, {Manchester}, \& {Wang}}]{ymw17}
{Yao}, J.~M., {Manchester}, R.~N., \& {Wang}, N. 2017, \apj, 835, 29,
  \dodoi{10.3847/1538-4357/835/1/29}

\bibitem[{{Zhang} {et~al.}(2021){Zhang}, {Wang}, {Wu}, {Wang}, {Li}, {Dai}, \&
  {Zhang}}]{zww+21}
{Zhang}, G.~Q., {Wang}, P., {Wu}, Q., {et~al.} 2021, \apjl, 920, L23,
  \dodoi{10.3847/2041-8213/ac2a3b}

\end{thebibliography}
\end{document}